\documentclass[preprint,3p]{elsarticle}
\usepackage{amsmath,amssymb}
\usepackage{xspace}
\usepackage{longtable}

\usepackage{ulem}
\usepackage{color}
\usepackage[mathscr]{eucal}
\usepackage{pifont}
\usepackage{listings}
\usepackage{slashed}
\usepackage{cancel}

\newcommand\SARAH{{\tt SARAH}\xspace}
\newcommand{\SARAHv}[1]{\SARAH~\texttt{#1}\xspace}
\newcommand\MG{{\tt MadGraph}\xspace}
\newcommand{\MGv}[1]{\MG~{\texttt{#1}}\xspace}
\newcommand\FeynArts{{\tt FeynArts}\xspace}
\newcommand\FormCalc{{\tt FormCalc}\xspace}
\newcommand\CalcHep{{\tt CalcHep}\xspace}
\newcommand\CompHep{{\tt CompHep}\xspace}
\newcommand\WHIZARD{{\tt WHIZARD}\xspace}
\newcommand\OMEGA{{\tt O'Mega}\xspace}
\newcommand\SPheno{{\tt SPheno}\xspace}
\newcommand\Vevacious{{\tt Vevacious}\xspace}
\newcommand\MO{{\tt MicrOmegas}\xspace}
\newcommand\HB{{\tt HiggsBounds}\xspace}
\newcommand\HS{{\tt HiggsSignals}\xspace}

\newcommand\UFO{{\tt UFO}\xspace}
\newcommand\pyrate{{\tt PyR@TE}\xspace}
\newcommand\Susyno{{\tt Susyno}\xspace}
\newcommand\Mathematica{{\tt Mathematica}\xspace}
\newcommand\Python{{\tt Python}\xspace}

\definecolor{mygray}{rgb}{0.3,0.3,0.3}

\title{\SARAHv{4}: A tool for (not only SUSY) model  builders}

\author{Florian Staub} 
\ead{fnstaub@th.physik.uni-bonn.de}
\address{Bethe Center for Theoretical Physics \& Physikalisches Institut der 
 Universit\"at Bonn, \\
Nu{\ss}allee 12, 53115 Bonn, Germany}

\begin{document}
\begin{flushright}
Bonn-TH-2013-17
\end{flushright}

\begin{abstract}
We present the new version of the \Mathematica package \SARAH which 
provides the same features for a non-supersymmetric model as previous
versions for supersymmetric models. This includes an easy and straightforward
definition of the model, the calculation of all vertices, mass matrices, 
tadpole equations, and self-energies. Also the two-loop renormalization 
group  equations for a general gauge theory are now included and 
have been validated with the independent \Python code \pyrate. 
Model files for \FeynArts, \CalcHep/\CompHep, \WHIZARD and in
the \UFO format can be written, and source code for \SPheno for the 
calculation of the mass spectrum, a set of precision 
observables, and the decay widths and 
branching ratios of all states can be generated. 
Furthermore, the new version includes routines to output  
model files for \Vevacious for both, supersymmetric and non-supersymmetric, models.
Global symmetries are also  
supported with this version and by linking \Susyno
the handling of Lie groups has been improved and extended. 
\end{abstract}

\maketitle

\section*{Program Summary}
{\bf Manuscript Title}: SARAH 4:  A tool for (not only SUSY) model  builders \\
{\bf Author}: Florian Staub \\
{\bf Program title}: SARAH \\
{\bf Programming language}: Mathematica \\
{\bf Computers for which the program has been designed}: All Mathematica is available for \\
{\bf Operating systems}: All Mathematica is available for \\
{\bf Keywords}: Model building, gauge theory, Renormalization group equations, Vacuum constraints \\
{\bf CPC Library Classification}: 11.1, 11.6 \\
{\bf Reasons for the new version}: New features in the definition of 
models and a full support of non-supersymmetric models. New output for Vevacious.\\
{\bf Solution method}: Non-supersymmetric models are supported by the new possibility to define not only chiral superfields but also component fields. The renormalization group equations (RGEs) for a non-supersymmetric models are calculated by using the generic formulae for a general quantum field theory. 
\\
{\bf Unusual features}:  Calculation of non-supersymmetric RGEs includes effects of kinetic mixing as well as gauge dependence of running vacuum expectation values. SARAH is the first tool which can automatically create model files for Vevacious. Fully automatized derivation of all terms in the Lagrangian which are fixed by gauge invariance.  \\
{\bf Restrictions}:  Only renormalizable terms in the Lagrangian are supported. No support of fields with spin 2 or 3/2.   \\
{\bf Summary of revisions}:  support of non-supersymmetric models; calculation of renormalization group equations 
for a general gauge theory; link to Susyno for handling of non-SU(N) gauge groups; support of global symmetries; 
output of model files for Vevacious; support of aligned VEVs; calculation of gauge dependent parts of RGEs for VEVs in running of supersymmetric and non-supersymmetric models.\\
{\bf Does the new version supersede the previous version?}:  Yes, the new version includes all known features of previous versions but provides also the new features mentioned above. \\
{\bf Running time}: Loading the Standard Model: 1.6s; calculation of all vertices: 11.8s; calculation of all RGEs: 130.2s ; 
output for Vevacious model files: 0.1s; output of model files in UFO format: 0.8s; output of model files for FeynArts: 0.1s; output of model files for CalcHep: 0.8s; output of model files for WHIZARD: 3.5s; writing of source code for SPheno: 34.5s. All times measured on Lenovo X220 with Intel(R) Core(TM) i7-2620M CPU @ 2.70GHz.

\section{Introduction}
Previous versions of \SARAH \cite{Staub:2008uz,Staub:2009bi,Staub:2010jh,Staub:2012pb} 
have been optimized for an easy, fast 
and exhaustive study of supersymmetric (SUSY) models beyond  the 
minimal supersymmetric standard model (MSSM). The range of models 
covered by \SARAH is very broad and it has 
been successfully used 
to study many different SUSY models: singlet extensions
with and without CP violation \cite{Ender:2011qh,Graf:2012hh,Kaminska:2013mya}, 
models with $R$-parity violation \cite{List:2013dga,Dreiner:2013jta}, different kinds of seesaw 
mechanisms \cite{Abada:2011mg,BhupalDev:2012ru,DeRomeri:2012qd}, 
models with extended gauge sectors at intermediate scales \cite{Esteves:2010si,Krauss:2013jva}
or the SUSY scale \cite{O'Leary:2011yq,Hirsch:2012kv}, models with
Dirac Gauginos \cite{Frugiuele:2012pe,Benakli:2012cy}, and even more
exotic ones \cite{Alves:2012fx}.

The emphasis in the new
version has been to push also non-supersymmetric (non-SUSY) models beyond the standard 
model (BSM) to the same level of automatization. 
For this purpose, the input format has been adjusted to support the definition of 
fermionic and scalar fields separately beside the definition of superfields 
only used so far. The calculation of vertices, 
mass matrices, tadpole equations and one-loop corrections to 1- and 2-points 
functions have always been performed in component fields and the 
corresponding routines can be used now for SUSY as well as non-SUSY models. 
In contrast, the generic formulae for the 
calculation of SUSY RGEs are not valid in the non-SUSY case. Therefore, 
new routines for the calculation of the two-loop $\beta$-functions in a general
quantum field theory have been implemented. 

For SUSY and non-SUSY models the calculated vertices can be used to write model files for
\CalcHep/\newline\CompHep \cite{Pukhov:2004ca,Boos:1994xb}, 
\FeynArts/\FormCalc \cite{Hahn:2000kx,Hahn:2009bf}, 
\WHIZARD/\OMEGA \cite{Kilian:2007gr,Moretti:2001zz} as well as in the 
\UFO format \cite{Degrande:2011ua} which can be handled for instance by 
\MGv{5} \cite{Alwall:2011uj}, {\tt GoSam} \cite{Cullen:2011ac} and {\tt Herwig++} \cite{Gieseke:2003hm,Gieseke:2006ga,Bellm:2013lba}. 
In addition, due to the output of
Fortran source code for \SPheno \cite{Porod:2003um,Porod:2011nf} 
the calculation of the one-loop corrected mass spectrum, branching ratios 
and decays widths of all states is possible. Also precision observables are calculated
and an easy link to \HB \cite{Bechtle:2008jh,Bechtle:2011sb} and \HS \cite{Bechtle:2013xfa} exists.  

Furthermore, global symmetries are supported in the new version and a link between \SARAH and
\Susyno \cite{Fonseca:2011sy} has been established. By using routines of \Susyno 
for the handling of Lie groups, \SARAH is no longer restricted to $U(1)$ and $SU(N)$ 
gauge groups and the non-fundamental irreducible representations (irreps) of unbroken 
gauge groups can  be treated in a more convenient way since it no longer necessary to 
express them as tensor products of the fundamental representation. 

Finally, \SARAHv{4} can also produce model files for \Vevacious \cite{Camargo-Molina:2013qva}. 
The combination \SARAH--\SPheno--\Vevacious provides the possibility to check 
the one-loop effective potential of a given model and parameter point
for the global minimum. In those checks 
the possibility of dangerous vacuum expectation values (VEVs) of charged and/or colored scalars 
can be taken into account 
and the life-time of meta-stable vacua can be calculated using 
{\tt Cosmotransitions} \cite{Wainwright:2011kj}.

This paper is organized as follows: the basics about the download and installation 
are shown in sec.~\ref{sec:download} and in sec.~\ref{sec:input} we present the changes in the 
input format of \SARAH which were necessary to support global symmetries, to link \Susyno 
and to handle non-SUSY fields in a convenient way. In sec.~\ref{sec:RGEs} we discuss the 
calculation of RGEs for a general quantum field theory. In sec.~\ref{sec:spheno}
we explain the different possibilities to calculate the mass spectrum and other observables 
with \SPheno and how to pass this information to Monte Carlo (MC) tools. In 
sec.~\ref{sec:vevacious} we introduce the interface to \Vevacious
before we conclude in sec.~\ref{sec:conclusion}. Other new features in \SARAHv{4} as well as 
an overview about the most important commands and existing models are given in the appendix. 

\section{Download and Installation}
\label{sec:download}
\SARAHv{4} has been successfully tested with \Mathematica versions 7 to 9, 
but the support of earlier versions of \Mathematica has stopped.
It can be downloaded from
\begin{center}
{\tt http://sarah.hepforg.org }
\end{center}
After downloading the tar-file, the package is extracted via
\begin{lstlisting}
tar -xf SARAH-4.0.0.tar.gz
\end{lstlisting}
We are going to assume in the following that \SARAH has been 
installed in the directory {\tt [\$SARAH-Directory]}. 

To load \SARAH, start \Mathematica and execute
\begin{lstlisting}
<<[$SARAH-Directory]/SARAH.m 
\end{lstlisting}
Afterwards, a model (called {\tt [\$MODEL]} in the following) which is
already included in \SARAH is initialized via
\begin{lstlisting}
Start["[$MODEL]"]; 
\end{lstlisting}
For instance, to load the MSSM {\tt [\$MODEL]}={\tt MSSM} is used. 
For an overview of commands which can be used to study the model after the initialization 
we refer to \ref{app:commands} and to the manual for more details. A list of all 
models delivered with \SARAHv{4} is given in \ref{app:models} and the implementation 
of new models was discussed in Refs.~\cite{Staub:2009bi,Staub:2010jh}.
We concentrate in the following on new features 
in \SARAHv{4}.

\section{New Possibilities in the Definition of a Model}
\label{sec:input}
\subsection{Global Symmetries}
\label{sec:global}
Beginning with \SARAHv{4} $Z_N$ and $U(1)$ global symmetries can be defined. 
For this purpose, a new array {\tt Global} has been introduced:
\lstset{frame=shadowbox}
\begin{lstlisting}
Global[[1]] = {Z[2], RParity};
Global[[2]] = {U[1], PecceiQuinn};
\end{lstlisting}
\lstset{frame=none}
First, the kind of the symmetry is defined and afterwards a name is given to the symmetry. 
In principle, up to 99 different global symmetries can be added for one model. 
The charges of all fields with respect to these groups are included in the definition 
of each field as discussed in sec.~\ref{sec:fields}. By convention, all $Z_N$ symmetries are 
always taken to be multiplicative symmetries. This means, the charges $Q$ of  additive symmetries 
have to be given as $\exp(2 i \pi Q)$. In addition, there is one special keyword to name a $U(1)$: 
if the symmetry is called {\tt RSymmetry} also the $R$ charges of the SUSY coordinates are included 
in the checks for charge conservation. See for example the minimal $R$-symmetric SSM (MRSSM)
which is included in \SARAH.\\

The global symmetries are taken into
account during the following operations:
\begin{enumerate}
 \item {\bf Checking the given (super)potential for charge conservation}: all terms in the potential 
 and superpotential defined by the user are checked since \SARAHv{2} for gauge invariance. In 
 the new version also the global symmetries are considered and if terms violate any of these symmetries
 a warning is printed. 
 \item {\bf Checking for all possible terms in the (super)potential}: starting with version 3 \SARAH 
 provides routines which
 check for all (renormalizable) terms in the superpotential for the given superfield content 
 which are in agreement with gauge invariance and $R$-parity. Now, all global symmetries are taken 
 into account during these checks. Also similar routines exist which check for all possible (renormalizable)
 terms in the   potential of a non-SUSY model. To perform these checks run {\tt CheckModel} after the initialization 
 of a model. 
 \item {\bf Adding SUSY soft-breaking masses for scalars}: \SARAH adds automatically for a SUSY model the 
 soft mass terms for all scalars. These are usually of the form $m^2 |\phi_i|^2$ where $i$ labels the different 
 scalars. In the case that fields have the same quantum number, also 
 terms of the form $m^2 \phi_i \phi^*_j + h.c.$  can exist in principle. So far, \SARAH checked only the gauge quantum 
 numbers and $R$-parity to include/exclude such terms. Now, all global symmetries are considered.
 \item {\bf Output for \MO}: the model files for \CalcHep which are supposed to be used with \MO
 must contain the information 
 about a conserved $Z_2$ symmetry. By default, this symmetry is now taken to be the one called {\tt RParity}
 in {\tt Global}. If no $Z_2$ symmetry with this name is exists, the first $Z_2$ symmetry defined in 
 {\tt Global} is used instead. 
\end{enumerate}
So far $R$-parity has been the only discrete symmetry which was supported by \SARAH. For this purpose, 
the $R$-parity of each particle has been given in the particle definitions in the file {\tt particles.m}. This 
entry has been become obsolete and is ignored now. In the same way, the flag {\tt RParityConservation = True/False}
which could be added to the model file has no longer any effect, but it is checked if a global symmetry called {\tt RParity}
is present or not. 

\subsection{Gauge Sector}
\label{sec:gauge}
The handling of Lie groups has been significantly improved by linking routines 
from the \Mathematica package {\tt Susyno} to get the representation matrices,
Clebsch-Gordan coefficients (CGCs) and gauge group constants. 
With doing this, not only $U(1)$ and $SU(N)$
groups can be handled, but also $SO(N)$, $Sp(2N)$ and expectational groups are supported.
Non-$SU(N)$ gauge groups are defined in the same way as $SU(N)$ ones:
\lstset{frame=shadowbox}
\begin{lstlisting}
Gauge[[1]]={  X,  U[1],   extra, g1,  False}; 
Gauge[[2]]={G10,  SO[10], SOGUT, g2,  False}; 
\end{lstlisting}
\lstset{frame=none}
First, the names of the gauge fields are given, then the kind of the group is defined. The third and fourth entry define the 
name of the gauge group and of the corresponding gauge coupling. In the last entry it is defined 
if the sums over charge indices of the gauge group should be expanded ({\tt True}) or not ({\tt False}). 
The last entry has no effect for Abelian groups. 

The names of the gauge fields are derived as follows 
from the first entry of each gauge group: gauginos start with {\tt f} (i.e. {\tt fX}, {\tt fG10}), 
the names of vector bosons are extended by {\tt V} as first letter (i.e. {\tt VX}, {\tt VG10}) 
and the corresponding ghosts start with a {\tt g}
(i.e. {\tt gX}, {\tt gG10}). In the case that the gauge groups carry a charge under a global symmetry, 
these are added at the end. Here, the same conventions are used as for matter fields discussed in 
the next section.

\subsection{Matter Fields}
\label{sec:fields}
\subsubsection{Superfields, Scalars and Fermions}
\SARAH has so far only supported chiral superfields in the matter sector 
which were defined in an array called {\tt Fields}. This has been a limitation for the study of non-SUSY fields
in contrast to other 
tools like {\tt LanHEP} \cite{Semenov:1998eb,Semenov:2002jw,Semenov:2008jy,Semenov:2010qt} 
or {\tt FeynRules} \cite{Christensen:2008py,Christensen:2010wz} which support the definition of component fields. 
This restriction is no longer present, but three possibilities to define matter fields exist now:
{\tt SuperFields}, {\tt FermionFields} and {\tt ScalarFields}. 
In this way it is now as simple to define a non-SUSY model in \SARAH as it has been 
for SUSY models all the time. For backward compatibility, {\tt Fields} is still supported by \SARAH but all 
model files delivered with \SARAH have been changed to the new format. 
We have to stress that the conventions of {\tt SuperFields} are slightly different to those of {\tt Fields}:
the first  and third entry have been exchanged. In addition, the 
transformation properties with respect to the global symmetries are added after the quantum 
numbers with respect to the gauge groups.

\paragraph*{Conventions to define global charges} If only one quantum number is given per superfield
per global symmetry this number is used for the superfield itself but also for the scalar 
and fermionic component. To define a symmetry which distinguishes between the superfield and 
its components, a list with three entries has to be given. 
For chiral superfields, the first entry is the charge for the superfield, the second for the scalar 
component, the third for the fermionic component. For vector superfield, the second entry refers 
to the gaugino, the third to the gauge boson. 
A convenient possibility might be to define new variables for such entries, e.g.
\lstset{frame=shadowbox}
\begin{lstlisting}
Global[[1]] = {Z[2], RParity};
RpM = {-1, -1, 1};
RpP = { 1,  1, -1};
\end{lstlisting}
with {\tt RpM} for $R$-parity -1 and {\tt RpP} for $R$-parity +1. \\
The charges under global symmetries defined for the gauge eigenstates are added during all rotations 
also to the mixed eigenstates as long as all fields in the basis of the rotation have the same charge. If 
the charges are different, the rotated fields don't get any charge under that global symmetry but a 
warning is printed. For example, if someone would have chosen the standard definition of $R$-parity 
but mixes sneutrinos and Higgs fields, the corresponding mass eigenstate wouldn't have a definite 
$R$-parity.

\paragraph*{Definition of fields} For a realistic model we define first the SM gauge groups 
using the conventions of sec.~\ref{sec:gauge} via
\begin{lstlisting}
Gauge[[1]]={B,   U[1], hypercharge, g1, False, RpM};
Gauge[[2]]={WB, SU[2], left,        g2, True,  RpM};
Gauge[[3]]={G,  SU[3], color,       g3, False, RpM}; 
\end{lstlisting}
Here, the last entries  in each line define the $R$-parity of the corresponding vector superfields and their components. 
The definition of the quark and Higgs doublets in the MSSM with respect to these gauge groups 
reads with the new 
conventions of \SARAHv{4}
\begin{lstlisting}
SuperFields[[1]] = {q,  3, {uL, qL},   1/6, 2, 3, RpM};
...
SuperFields[[3]] = {Hd, 1, {Hd0, Hdm}, 1/2, 2, 1, RpP};
\end{lstlisting}
First, a name for the superfield is given, in the second entry the number of generations is set and in the third entry 
the names for 
the isospin components are defined. Afterwards, the hypercharge, and the transformation under $SU(2)_L \times SU(3)_C$ 
are given, the last entry states the $R$-parity using the variables {\tt RpM} and {\tt RpP} defined above. 
The syntax for scalar and fermion fields in the case  of
non-SUSY models looks exactly the same:
\begin{lstlisting}
FermionFields[[1]] = {q, 3, {uL, qL}, 1/6, 2, 3};
...
ScalarFields[[1]] =  {H, 1, {H0, Hm}, 1/2, 2, 1};
\end{lstlisting}
Here, we have not assumed any global symmetry. Thus, the definition of the gauge groups is the same as 
for the MSSM above just without the last entry in each line. 
By default all scalars are taken to be complex. To define them as real the name of the field has to 
be added to the list {\tt RealScalars}. For instance, a real gauge singlet is added to the model by
\begin{lstlisting}
ScalarFields[[2]] = {S, 1, s,  0, 1, 1};
RealScalars = {S};
\end{lstlisting}
Note, the fermionic components of 
superfields start with {\tt F} and the scalars with {\tt S}, 
i.e. the quark doublets stemming from the superfield definition above are called {\tt FuL, FdL} and 
the squarks are called {\tt SuL, SdL}. For fermion and scalar fields defined separately, no 
renaming takes place but the fields are used exactly as they are defined. 
It is not possible to use {\tt SuperFields} and {\tt FermionFields} or {\tt ScalarFields} at the same time. 
If the user wants to define scalars or fermions, all superfields have to be written as components.

\subsubsection{Handling of non-fundamental Representations}
The link to \Susyno has opened a new possibility to handle non-fundamental 
representations of non-Abelian gauge groups: 
so far these fields have been always treated as tensor products. While
this is still done for $SU(N)$ gauge groups which get broken, 
irreps of unbroken gauge groups are written as vectors with the appropriate dimension. 
This has especially advantages for models which contain color sextets: for these 
models it is possible now to write model files in the {\tt UFO} format which can be used with 
\MGv{5}. The same would hold for all other
color representations like decuplets, but those are not yet supported by any MC tool. 

To treat non-fundamental irreps as vectors under the corresponding gauge group, a major 
rewriting of core routines of \SARAH has been necessary. This has several consequences:
\begin{itemize}
 \item The list of all existing particles does not only contain the indices of 
 each particle, but also the dimension of each index is explicitly given. For instance, 
 the entry of the d-squarks in {\tt Particles[EWSB]} changed from
 \begin{lstlisting}
   {Sd, 1, 6, S, {generation, color}}
 \end{lstlisting}
 to
 \begin{lstlisting}
   {Sd, 1, 6, S, {{generation, 6}, {color, 3}}}
 \end{lstlisting}
\item There is no longer a special name for charge indices of the adjoint representation (so far {\tt a <> index}), 
i.e.  the entry of the gluon in {\tt Particles[EWSB]} has changed from
\begin{lstlisting}
{VG, 1, 1, V, {acolor,lorentz}}
\end{lstlisting}
to
\begin{lstlisting}
{VG, 1, 1, V, {{color, 8}, {lorentz, 4}}}
\end{lstlisting}
\item There is just one charge index per particle per unbroken gauge group, 
i.e. non-fundamental irreps carry less indices than before. For instance, 
a fermionic color sextet which was so far treated as $3\times 3$ matrix with two color indices of dimension 3 
\begin{lstlisting}
{Fsix, 1, 1, F, {color, colorb}}
\end{lstlisting}
is now defined internally as
\begin{lstlisting}
{Fsix, 1, 1, F, {{color, 6}}}
\end{lstlisting}
\end{itemize}

\subsubsection{Defining Matter Interactions}
The syntax to define the superpotential has been simplified. It is no longer necessary to give it 
as a list of terms but a more intuitive form as sum is possible. For instance, instead of the former definition of the superpotential
of the MSSM 
\begin{lstlisting}
SuperPotential = { {{1, Yu},{u,q,Hu}}, {{-1,Yd},{d,q,Hd}},
                   {{-1,Ye},{e,l,Hd}}, {{1,\[Mu]},{Hu,Hd}}}; 
\end{lstlisting}
the short form
\begin{lstlisting}
SuperPotential = Yu u.q.Hu - Yd d.q.Hd - Ye e.l.Hd + \[Mu] Hu.Hd; 
\end{lstlisting}
can be used.

In the same way, the potential for non-SUSY models is defined as sum of terms
\begin{lstlisting}
LagNoHC = -Mu2 conj[H].H + 1/2 \[Lambda] conj[H].H.conj[H].H;
LagHC =  Yd conj[H].d.q + Ye conj[H].e.l + Yu H.u.q;

DEFINITION[GaugeES][LagrangianInput]= {
	{LagHC,  {AddHC->True}},
	{LagNoHC,{AddHC->False}}
};
\end{lstlisting}
Here, we defined all interactions in terms of gauge eigenstates ({\tt GaugeES}) and all 
rotations to the mass eigenstates are performed by \SARAH. In addition, there exists the option that \SARAH adds
automatically the Hermitian conjugated for the given part of the Lagrangian ({\tt AddHC->True}).

It is neither necessary for the superpotential nor for terms added via {\tt DEFINITION[\_][LagrangianInput]} to define 
any index structure. This task is automatically performed by \SARAH: it adds generation indices if necessary 
to all fields and parameters and contracts 
all charges using either the Kronecker delta ({\tt Delta}), the antisymmetric tensor ({\tt epsTensor})
or CGCs. CGCs are a new feature in \SARAHv{4} since in previous versions 
it has been sufficient just to use the Kronecker delta and antisymmetric tensor 
because of the treatment of all irreps as tensor product. The CGCs which are now used for interactions involving 
higher dimensional irreps are parametrized as follows:
\begin{lstlisting}
CG[group,dnykin][indices]
\end{lstlisting}
First, the gauge group is given, afterwards the Dynkin labels of all involved irreps are stated and finally
the indices are listed. For instance, the interaction between a scalar color sextet ({\tt six}) and two fermionic triplets 
which appear in several generations  ({\tt t}) is defined in the input by the user just as
\begin{lstlisting}
Ys six.t.t
\end{lstlisting}
and internally expanded by \SARAH to
\begin{lstlisting}
Ys[gen2,gen3]*CG[SU[3],{{0,2},{1,0},{1,0}}][col1,col2,co3]*
                            six[{col1}]*t[{gen2,col2}]*t[{gen3,col3}] 
\end{lstlisting}
The numerical values for all CGCs are calculated by \Susyno. The user can check the array
\lstset{frame=none}
\begin{lstlisting}
SA`LagrangianContractions 
\end{lstlisting}
\lstset{frame=shadowbox}
to see what contractions have been used by \SARAH for each term. \\

\SARAH takes by default the first gauge invariant contraction of indices it finds. That's usually the simplest contractions which 
exists. That means that always co- and contravariant indices are contracted which are next to each other, e.g.
$\phi^{ij}\Psi_k \Psi_l \delta_{jk} \delta_{il}$ is used but not $\phi^{ij}\Psi_k \Psi_l \delta_{ik} \delta_{jl}$. 
Furthermore, it is also assumed that the contraction is not a sum of different products involving Kroneckers deltas 
and antisymmetric tensors but just one term. For instance, $\phi^i \phi_j \phi_k \phi^l$ is contracted 
with $\delta_{ij} \delta_{kl}$ while another gauge invariant form would be the result of the completeness 
relation of $SU(N)$ gauge groups: $2 \delta_{il} \delta_{jk} - \frac{2}{N} \delta_{il} \delta_{jk}$. 
Since several possibilities to contract the indices of one term might exist, it is also possible that the user can define a contraction for each term which is then used by \SARAH. 
\paragraph*{Example} It might for instance be necessary for (renormalizable) interactions
which are an effect of fields integrated out to define the contractions by hand. 
For instance, D-terms of heavy vector superfields can give rise to
$|H_u^\dagger \sigma_a H_u + H_d^\dagger \sigma_a H_d|^2$ \cite{Bharucha:2013ela}. 
This can be written by making use of the 
completeness relation of the Pauli matrices as
\begin{lstlisting}
(2 Delta[lef1,lef4] Delta[lef2,lef3]- Delta[lef1,lef2] Delta[lef3,lef4])*
     (conj[SHd].SHd.SHd.conj[SHd] +  conj[SHu].SHu.SHu.conj[SHu] 
                                      - 2 conj[SHd].SHd.SHu.conj[SHu]) ) 
\end{lstlisting}
\smallskip\smallskip

\paragraph*{Symmetry of parameters} It is also new that parameters which involve several times the same fields which come in more than one generation
are tested for their symmetry: \SARAH checks if such a parameter is symmetric, anti-symmetric, hermitian, or anti-hermitian 
and simplifies expression by using these properties during the different calculations. For instance, {\tt Ys} introduced above would be defined as symmetric while it would be anti-symmetric
if the color triplets would also be doublets under $SU(2)$. The entry {\tt Symmetry} in {\tt parameters.m}
which has been used in earlier versions to manually define these symmetries is no longer necessary and ignored now.

\subsection{The Lagrangian derived by \SARAH}
\SARAH derives from the definition of the gauge and matter sector and the (super)potential the full Lagrangian of the model. 
For this purpose it generates automatically all interactions which are fixed by gauge and, if considered, SUSY invariance, i.e. kinetic terms, 
self-interactions of vector bosons, D-terms and gaugino interactions. From the kinetic terms also the gauge fixing terms in $R_\xi$ gauge are derived and the ghost interactions are calculated. In the case of SUSY models the standard soft terms are added by default, too. Because of this level of automatization the input files are usually very short, they can easily be extended and the possibilities for the user to make errors are reduced. 
The main restriction to define interactions at the moment is that only renormalizable terms are fully supported. For SUSY models it is possible to define superpotential terms with four superfields. However, only the dimension 5 operators (which enter, for instance, the fermion masses) are derived from that. For the potential itself it is only possible to add renormalizable terms for now. \\

Examples for the full model files of a SUSY and non-SUSY model in the new format are given in \ref{app:NMSSM} (NMSSM) and \ref{app:SM} (Standard Model).

\section{Renormalization Group Equations for a general Quantum Field Theory}
\label{sec:RGEs}
\subsection{Non-SUSY RGEs with \SARAH}
Since version 2 \SARAH calculates the full two-loop RGEs for a SUSY theory based on the 
generic results given in Ref.~\cite{Martin:1993zk}. These calculations have been later generalized 
to the case of several Abelian gauge groups with full  kinetic mixing 
by using the substitution rules presented in Ref.~\cite{Fonseca:2011vn}. 
In version 3.2 the two-loop RGEs for
Dirac gauginos based on Ref.~\cite{Goodsell:2012fm} have been included. 
All of these calculations take into account the most 
general CP and flavor structure. The gauge dependence of running vacuum expectation values (VEVs)
is also included, see \ref{app:RunningVEVs}.\\

\SARAHv{4} calculates the RGEs for a general quantum field theory with the same 
precision as it is done for SUSY models based on the generic results of
Refs.~\cite{Machacek:1983tz,Machacek:1983fi,Machacek:1984zw,Luo:2002ti}.
Also the impact of kinetic mixing for non-SUSY models has been included by using the 
corresponding rules which became recently available in Ref.~\cite{Fonseca:2013bua}. 
Finally, the gauge dependence in the running of the VEVs is 
now included by using the results of Ref.~\cite{Sperling:2013eva}.  \\

The calculation of the RGEs can be started after the initialization of a model via
\lstset{frame=none}
\begin{lstlisting}
CalcRGEs[Options];
\end{lstlisting}
This is the same command which is also used to get the SUSY RGEs, i.e.  \SARAH checks which 
calculation is valid for the given model. The different options are
\begin{itemize}
\item \verb"TwoLoop", Value: \verb"True" or \verb"False", Default: \verb"True" \\
Defines, if also the two-loop RGEs should be calculated.
\item \verb"ReadLists", Value: \verb"True" or \verb"False", Default: \verb"False" \\
Defines, if previously calculated RGEs should be loaded. 
\item \verb"VariableGenerations", Value: List of particles, Default: \verb"{}"\\
Defines, that the generations of specific fields should be treated as variable. For this purpose variables 
\verb"nG[X]" are introduced, where \verb"X" is the name of the superfield (SUSY model) or fermion/scalar (non-SUSY model).
\item \verb"NoMatrixMultiplication", Values: \verb"True" or \verb"False", Default: \verb"False"\\
This can be used to switch on/off matrix multiplication in the simplification of the RGEs. 
\item \verb"IgnoreAt2Loop", Values: a list of parameters, Default: \verb"{}"\\
Can be used to define parameters which should be neglected during the calculation of the two-loop $\beta$-functions.
 \item \verb"WriteFunctionsToRun", \verb"True" or \verb"False", Default: \verb"True" \\
 Defines, if a file should be written to evaluate the RGEs numerically in \Mathematica, see  sec.~\ref{sec:RGErunning}
\end{itemize}
The results are saved in three dimensional arrays: the first entry is the name of  the considered parameter, the second entry is the one-loop \(\beta\)-function and the third one is the two-loop \(\beta\)-function. 
For non-SUSY models the different generic structures are saved in 
 \begin{itemize}
  \item  \verb"Gij": Anomalous dimensions of all fermions and scalars
  \item \verb"BetaGauge": Beta functions of all gauge couplings
  \item \verb"BetaLijkl": Beta functions of all  quartic scalar couplings
  \item \verb"BetaYijk": Beta functions of all  interactions between two fermions and one scalar
  \item \verb"BetaTijk": Beta functions of all  cubic scalar interactions
  \item \verb"BetaMuij": Beta functions of all  bilinear fermion terms
  \item \verb"BetaBij": Beta functions of all  bilinear scalar terms
  \item \verb"BetaVEV": Beta functions of all  VEVs
\end{itemize}
\paragraph*{Examples} Some entries for the Standard Model (SM) look as follows:
\begin{itemize}
 \item Hypercharge coupling  ({\tt BetaGauge[[1]]})
 \begin{lstlisting}
  {g1, 
  (41*g1^3)/10, 
  (g1^3*(199*g1^2 + 135*g2^2 + 440*g3^2 - 25*trace[Yd, Adj[Yd]] - 
   75*trace[Ye, Adj[Ye]] - 85*trace[Yu, Adj[Yu]]))/ 50}
 \end{lstlisting}
 \item $\lambda$ coupling ({\tt BetaLijkl[[1]]}; only the one-loop part is shown)
 \begin{lstlisting}
 {\[Lambda], 
   (-27*g1^4)/100 - (9*g1^2*g2^2)/10 - (9*g2^4)/4 - 
   (9*g1^2*\[Lambda])/5 - 9*g2^2*\[Lambda] - 12*\[Lambda]^2 + 
   12*\[Lambda]*trace[Yd, Adj[Yd]] + 4*\[Lambda]*trace[Ye, Adj[Ye]] + 
   12*\[Lambda]*trace[Yu, Adj[Yu]] + 12*trace[Yd, Adj[Yd], Yd, Adj[Yd]] + 
   4*trace[Ye, Adj[Ye], Ye, Adj[Ye]] + 12*trace[Yu, Adj[Yu], Yu, Adj[Yu]],  
   ...
  };
 \end{lstlisting}
\item Electron Yukawa coupling ({\tt BetaYijk[[1]]}; only the one-loop part is shown)
\begin{lstlisting}
{Ye[i1, i2], 
  ((-9*g1^2)/4 - (9*g2^2)/4 + 3*trace[Yd, Adj[Yd]] + 
   trace[Ye, Adj[Ye]] + 3*trace[Yu, Adj[Yu]])*Ye[i1, i2] 
   + (3*MatMul[Ye, Adj[Ye], Ye][i1, i2])/2,
  ... }
\end{lstlisting}
\item If the user wants to see the impact on the number of generations of quarks on the running gauge couplings, it is possible 
to run {\tt CalcRGEs[VariableGenerations->\{d,q,u\}]}. The $\beta$-functions involve then variables {\tt nG[X]} which correspond 
to the chosen number of generations for those particles, e.g. {\tt BetaGauge} reads
\begin{lstlisting}
{{g1, (g1^3*(57 + 4*nG[d] + 2*nG[q] + 16*nG[u]))/30, ...}, 
 {g2, (g2^3*(-37 + 6*nG[q]))/6, ...}, 
 {g3, (g3^3*(-33 + nG[d] + 2*nG[q] + nG[u]))/3, ...}} 
\end{lstlisting}

\end{itemize}

\subsection{Validation}
In literature the full two-loop RGEs for a non-SUSY model beyond the SM are 
hardly available. Especially, parameters with dimension of mass like cubic scalar interactions 
are usually not calculated with this precision. This has been one of the reasons to 
develop a completely independent code in {\tt python}. The result is the public 
package \pyrate \cite{Lyonnet:2013dna} which has been used to 
cross check the results obtained by \SARAH. We found full agreement between the 
results by \SARAH and \pyrate. In addition, the results have been checked against 
the literature for those models and terms for which a reference was available. 
We found widely agreement 
with the literature but could also spot some mistakes in the references. Details about all cross 
checks are given in Ref.~\cite{Lyonnet:2013dna}.

\subsection{Numerical Evaluation of the RGEs with \Mathematica}
\label{sec:RGErunning}
\SARAHv{4} does not only save the calculated RGEs in the internal format in separated
text files in {\tt [\$SARAH-Directory]/Output/[\$MODEL]/RGEs}, but writes in addition a file 
which can be used for a numerical evaluation of the SUSY and non-SUSY RGEs in \Mathematica. This file is stored 
in the same directory as the other results for the RGEs and is called {\tt RunRGEs.m}. It provides all RGEs in a format which 
is understood by \Mathematica and a function called {\tt RunRGEs} for the numerical evaluation based on 
{\tt NDSolve}.  The syntax of this function is as follows:
\begin{lstlisting}
RunRGEs[initialization, log(start), log(end), Options]; 
\end{lstlisting}
First, a list with the non-zero values of parameters are given. The second 
and third entry define the logarithm of the scale where the running starts and stops. 
The option defines if two-loop contributions should 
be included or not {\tt TwoLoop -> True/False}. By default the two-loop parts are 
taken into account. {\tt RunRGEs} returns the {\tt InterpolationFunction} obtained by
{\tt NDSolve}. 

\paragraph*{Example} A one-loop running of the gauge couplings from 1000~GeV to $10^{16}$~GeV is 
performed after loading {\tt RunRGEs.m} by
\begin{lstlisting}
<< "[\$SARAH-Directory]/Output/[\$MODEL]/RGEs/RunRGEs.m";
solution = RunRGEs[{g1->0.45, g2->0.63, g3->1.04}, 3,16, TwoLoop->False];
\end{lstlisting}
Here, we stored the output of {\tt RunRGEs} as variable, which can be used as follows
\begin{lstlisting}
{g1[16], g2[16], g3[16]} /. solution[[1]]; 
Plot[g1[x]/. solution[[1]], {x,3,16}];
\end{lstlisting}
\lstset{frame=none}
In the first line, the numerical values of all gauge couplings at $10^{16}$~GeV are shown, 
in the second line, the running of $g_1$ between $10^3$ and $10^{16}$~GeV is plotted.

\section{Mass Spectrum Calculation of non-SUSY Models with \SPheno and Interface to Monte Carlo Tools}
\label{sec:spheno}
\subsection{\SPheno and non-SUSY Models}
\SARAHv{3} has been the first `spectrum--generator--generator': using the derived information about the 
mass matrices, tadpole equations, vertices, loop corrections and RGEs for the given model \SARAH writes Fortran source code
for \SPheno. 
Based on this code the user gets a fully functional spectrum generator for the given model. 
The features of a spectrum generator created in this way are (i) a precise mass spectrum calculation using two-loop 
RGEs and full one-loop corrections to all masses, (ii) the calculation of branching ratios and decay width of all 
SUSY and Higgs fields as well as (iii) a prediction for precision observables like 
$b\to s\gamma$, $g-2$ and $B_{s,d}^0 \to l_i \bar{l}_j$ \cite{Dreiner:2012dh}. In addition, \SPheno modules created 
by \SARAH can write input files for \HB
and \HS. 

All of these possibilities are now also available for non-SUSY models as well. For non-SUSY models a 
few adjustments have happened. 
\begin{enumerate}
\item \SPheno calculates the SM gauge and Yukawa couplings 
 from the experimental data. The corresponding routines written by \SARAH assumed so far 
that two Higgs VEVs  ($v_d$,$v_u$) are presented. In \SARAHv{4} these routines got adjusted to 
support also models with just one Higgs VEV like the SM or simple extensions of it. 
\item In the case of a non-SUSY model, the translation 
from $\overline{\text{MS}}$ to $\overline{\text{DR}}$ parameters as well as the calculation of SUSY thresholds is skipped. 
\item Finally, all self-energies are calculated in $\overline{\text{MS}}$ instead of $\overline{\text{DR}}$ by using the new flexibility of the loop calculations 
shown in \ref{sec:loops}. 
\end{enumerate}

To create the \SPheno output for a given model run in \Mathematica
\begin{lstlisting}
<<[$SARAH-Directory]/SARAH.m;
Start["[$MODEL]"];
MakeSPheno[];
\end{lstlisting}
The source code for \SPheno will be stored in {\tt [\$SARAH-Directory]/Output/[\$MODEL]/EWSB/SPheno/}. 
To compile this code enter the directory of your \SPheno installation (which we are going to call 
{\tt [\$SPheno-Directory]}), create a new sub-directory and copy the code into this directory:
\begin{lstlisting}
cd [$SPheno-Directory]
mkdir [$MODEL]
cp [$SARAH-Directory]/Output/[$MODEL]/EWSB/SPheno/* [$MODEL]/
\end{lstlisting}
Afterwards, the code is compiled via 
\begin{lstlisting}
make Model=[$MODEL] 
\end{lstlisting}
and a new executable {\tt SPheno[\$MODEL]} is created. To run this executable 
an input file {\tt LesHouches.in.[\$MODEL]} is needed. 
\begin{lstlisting}
./bin/SPheno[$MODEL] [$MODEL]/LesHouches.in.[$MODEL] 
\end{lstlisting}
The entire output including all parameters, masses, branching ratios and low-energy 
observables is saved in {\tt SPheno.spc.[\$MODEL]}. 

\subsection{Setting the Properties of the \SPheno Version}
To create \SPheno versions with \SARAH all features of the \SPheno code are defined by an additional input file 
{\tt SPheno.m} of \SARAH. The content of this file for non-SUSY models is very similar to the one 
of SUSY models which has been discussed in great detail in Ref.~\cite{Staub:2011dp}. We summarize here the main 
aspects. 

In general, there are two different kinds of \SPheno versions the user can create: a 'low scale' version which expects 
all parameters at the renormalization scale, or a 'high scale' version which includes the possibility to 
perform a RGE running to some higher scale like the GUT or a threshold scale. We are going to use the synonym 'GUT scale' in the following 
even if this scale must not necessarily be connected to a grand unified theory and can even be much lower than $10^{16}$~GeV. 
For both cases {\tt SPheno.m} looks slightly different 
and we are going to discuss both options. 

\subsubsection{'Low Scale' \SPheno Version}
The flag to create a low scale \SPheno version is {\tt OnlyLowEnergySPheno = True}.
In addition, the list of input parameters ({\tt MINPAR}), the 
boundary conditions at the input scale ({\tt BoundaryLowScaleInput}), 
the parameters fixed by the tadpole equations ({\tt ParametersToSolveTadpoles})
and the particles for which 2- and 3-body  decays are calculated  ({\tt ListDecayParticles}, {\tt ListDecayParticles3B}) 
is needed. Altogether, {\tt SPheno.m} to get a low scale version of the SM looks like
\lstset{frame=shadowbox}
\begin{lstlisting}
OnlyLowEnergySPheno = True;

MINPAR={{1,LambdaIN}};

ParametersToSolveTadpoles = {Mu2};

BoundaryLowScaleInput={
 {\[Lambda],LambdaIN}
};

ListDecayParticles = {Fu,Fe,Fd,hh};
ListDecayParticles3B = {{Fu,"Fu.f90"},{Fe,"Fe.f90"},{Fd,"Fd.f90"}};
\end{lstlisting}

\subsubsection{'High Scale' \SPheno Version}
For a \SPheno version which includes the RGE running up to a higher scale, 
{\tt OnlyLowEnergySPheno = False} is set. Other differences in comparison to the 
low scale version is the definition of the renormalization scale
({\tt RenormalizationScaleFirstGuess}, {\tt RenormalizationScale}) as well as the boundary 
conditions at three different scales: (i) at $M_Z$ ({\tt BoundaryEWSBScale}), 
(ii) at the renormalization scale  ({\tt BoundaryRenScale}), (iii) at the GUT 
scale ({\tt BoundaryHighScale}). Here 
{\tt BoundaryRenScale} is the equivalent to {\tt BoundarySUSYScale} for SUSY models. 
{\tt RenormalizationScaleFirstGuess} is the renormalization scale used in the first iteration, while 
{\tt RenormalizationScale} is used in all following iterations. The main difference is that 
{\tt RenormalizationScale} can depend on parameters or masses which are calculated by \SPheno while 
{\tt RenormalizationScaleFirstGuess} must be a constant or completely fixed by the input. 
In addition, using the variable {\tt ConditionGUTscale} it is possible to define a condition for the GUT scale,
i.e. what relation among parameters should be fulfilled to stop the RGE running and to apply the boundary conditions 
given in {\tt BoundaryHighScale}. 

With these conventions {\tt SPheno.m} for the SM including RGE running reads
\begin{lstlisting}
OnlyLowEnergySPheno = False;

MINPAR={{1,LambdaIN}};

ParametersToSolveTadpoles = {Mu2};

RenormalizationScaleFirstGuess = 100^2;
RenormalizationScale = v^2;

ConditionGUTscale = g1 == g2;

BoundaryHighScale = { {\[Lambda],LambdaIN} }; 
BoundaryEWSBScale = {}; 
BoundaryRenScale={ };

ListDecayParticles = {Fu,Fe,Fd,hh};
ListDecayParticles3B = {{Fu,"Fu.f90"},{Fe,"Fe.f90"},{Fd,"Fd.f90"}};
\end{lstlisting}
Here we put as condition to stop the RGE running that the running values of 
$g_1$ and $g_2$  should be equal. 
However, in contrast to SUSY models it is not very common to run 
the RGEs up to the scale where two gauge couplings do meet. Often 
just the running parameters at some specific scale should be extracted or 
boundary conditions at some threshold scale where one assumes the theory to break 
down should be applied. 
For this purpose, another possibility exists to define the scale where the 
running stops: it is possible to set the 'GUT scale' to 
a constant value in the Les Houches input file
\begin{lstlisting}
Block SPhenoInput    # SPheno specific input 
....
31 2.000E+16         # fixed GUT scale (-1: dynamical GUT scale)  
\end{lstlisting}
while with 
\begin{lstlisting}
Block SPhenoInput    # SPheno specific input 
....
31 -1                # fixed GUT scale (-1: dynamical GUT scale)  
\end{lstlisting}
the condition defined by {\tt ConditionGUTscale} is used to get a dynamical GUT scale.

\subsection{Interfacing \SPheno and Monte Carlo Tools}
The information obtained by \SPheno can be directly used together with the corresponding \SARAH 
model files for \CalcHep, \MG or \WHIZARD  to make Monte Carlo (MC) studies. We show 
in the following how the information 
can be exchanged between \SPheno and the different tools.

\subsubsection{\CalcHep}
In the case of \CalcHep it is necessary to make sure that \SARAH writes the corresponding model files 
using the {\tt SLHA+} functionality of \CalcHep \cite{Belanger:2010st} which enables \CalcHep to read SLHA files:
\lstset{frame=none}
\begin{lstlisting}
MakeCHep[SLHAinput -> True]; 
\end{lstlisting}
This is also the default option of {\tt MakeCHep[]} and it is used if the user hasn't changed the default settings before. In this case, \CalcHep extracts from the \SPheno spectrum 
file the information about all parameters, masses and rotation matrices. 
For this purpose the spectrum file must be located 
in the directory in which {\tt n\_calchep} is executed. Of course, if the \SARAH model files for \CalcHep are used 
together with \MO, it is also sufficient to copy the \SPheno spectrum file to the directory of the \MO main file which is used for the dark matter study. 

\subsubsection{\MG}
To use the information obtained by \SPheno together with \MG, replace the file {\tt param\_card.dat} in {\tt [\$MadGraph-Directory]/[\$Project]/Cards/} 
by the \SPheno spectrum file. Here, {\tt [\$Project]} is the \MG subdirectory which has been generated for the current MC study.

\MG has problems with reading the \HB specific blocks in the \SPheno spectrum file ({\tt HiggsBoundsInputHiggsCouplingsFermions} and 
{\tt HiggsBoundsInputHiggsCouplingsBosons}). Therefore, either the output of this information must be disabled by
setting 
\lstset{frame=shadowbox}
\begin{lstlisting}
Block SPhenoInput   # SPheno specific input 
...
520 0.              # Write effective Higgs couplings (HiggsBounds blocks) 
\end{lstlisting}
in the Les Houches input file of \SPheno, or the blocks must be deleted before running \MG. 

There is another subtlety: \SARAH initializes by default all unknown parameters with 0 in the {\tt UFO} file. Therefore, it can happen that 
\MG complains about a division by 0 when it generates the output  and calculates internal parameters before loading a spectrum file. If this happens the user should put a non zero initialization value in {\tt parameters.py} for the corresponding parameter.

\subsubsection{\WHIZARD}
To link \WHIZARD and \SPheno, all \SPheno modules created by \SARAH write the information about the parameter and masses into an additional file 
called {\tt WHIZARD.par.[\$Model]} in the \WHIZARD specific format. To get this output, the following flag has to be set in the Les Houches input file of \SPheno
\begin{lstlisting}
Block SPhenoInput   # SPheno specific input 
...
75 1                # Write WHIZARD files 
\end{lstlisting}
The parameter file can then be included in the Sindarin input file for \WHIZARD  via
\begin{lstlisting}
include("[$SPheno-Directory]/WHIZARD.par.[$MODEL]") 
\end{lstlisting}

\section{\Vevacious}
\label{sec:vevacious}
\Vevacious \cite{Camargo-Molina:2013qva} is a new tool to check for the global minimum of the 
one-loop effective potential for a
given model allowing for a particular set of non-zero VEVs. 
The importance of these checks has 
recently be demonstrated at the example of color and charge breaking minima in the MSSM \cite{Camargo-Molina:2013sta}. 
As input \Vevacious needs the tadpole equations, the polynomial part of the potential and all mass 
matrices for the model assuming that all allowed VEVs are non-zero. This means that, for instance, 
if stau VEVs should be studied a common basis of all uncolored scalars and fermions has to be defined
and also the neutral gauge boson will mix with the $W$ bosons. Usually, these modifications of the \SARAH 
model file are easily done and examples for the MSSM with different charge and color breaking scenarios 
are included in {\tt [\$SAHRAH-Directory]/Models/CCB-MSSM}. For a detailed discussion 
of the input format of \Vevacious we refer to Ref.~\cite{Camargo-Molina:2013qva}. 

The model files for \Vevacious for a model are generated after loading \SARAH and initializing the
model by the command {\tt MakeVevacious[]}:
\lstset{frame=none}
\begin{lstlisting}
<<[$SARAH-Directory]/SARAH.m;
Start["[$MODEL]"];
MakeVevacious[Options];
\end{lstlisting}
The following options are possible
\begin{itemize}
\item {\tt ComplexParameters}, Value: list of parameters, Default: \{\}: \\
 By default, all parameters are assumed to be real when writing the \Vevacious
 input files. However, the user can define those parameters which should be
 treated as complex. 
 \item {\tt IgnoreParameters}, Value: list of parameters, Default: \{\}: \\
 The user can define a list of parameters which should be set to zero when
 writing the \Vevacious input.
 \item {\tt OutputFile}, Value: String, Default: {\tt [\$Model].vin}: \\
 The name used for the output file. 
 \item {\tt Scheme}, Value: {\tt DR} or {\tt MS}, Default:  {\tt DR} for SUSY models, {\tt MS} for non-SUSY models: \\
 Defines, if as renormalization scheme $\overline{\text{DR}}'$ or $\overline{\text{MS}}$ should be used. See for details 
 of the effective potential in the different schemes Ref.~\cite{Martin:2001vx}.
 \end{itemize}
The first two options allow to treat parameters differently in the
 \Vevacious output as defined in the \SARAH model file.
 The reason is that the evaluation of a parameter point with \Vevacious 
 can be very time consuming and subdominant terms can be neglected
 using these options. That might speed up the run \Vevacious, what  
 is however not yet quantified.

\section{Conclusion}
\label{sec:conclusion}
We have presented the new features in \SARAHv{4}. The focus has been on a  largely improved and extended support of non-SUSY models. In particular, the complete two-loop RGEs for a general quantum field theory have been implemented.  We have shown how \SPheno can be used to calculate the mass spectrum and other observables for non-SUSY models and how this information is passed to MC tools. 
By linking \Susyno, the handling of Lie groups has been generalized and models with color sextets can be treated in a way suitable for MC tools. Also the definition of global symmetries is now possible. Finally, the interface to \Vevacious allows to perform checks 
for the global minimum in a huge variety of SUSY and non-SUSY models. 

\section*{Acknowledgments}
We are in debt to Renato Fonseca for providing us to the powerful routines of \Susyno for the handling of Lie groups.
We thank Daniel Busbridge, Eliel Camargo, Philip Diessner,  Mark Goodsell,  Wojciech Kotlarski,  Manuel Krauss,  Florian Lyonnet,  Moritz McGarrie, Kilian Nickel, Ben O'Leary, Werner Porod, Ingo Schienbein, Kai Schmidt-Hoberg, Ken Van Tilburg, Alexander Voigt, and Akin Wingerter 
for their feedback, fruitful discussions and helpful suggestions.

\begin{appendix}
\section{Other Improvements in \SARAHv{4}}
\subsection{Gauge-Gravity Anomalies}
Since version 2 \SARAH checks for all triangle gauge anomalies as well as Witten anomalies
for the given particle content in the considered model. 
These checks have now be extended and also gauge-gravity anomalies are included. 

\subsection{Definition of complex and aligned Vacuum Expectation Values}
The general definition of the decomposition of a complex scalar field into its real
components and a VEV reads in \SARAH
\lstset{frame=shadowbox}
\begin{lstlisting}
DEFINITION[$EIGENSTATES][VEVs]= 
  {{Scalar Field, {$VEV, coefficient}, {$PSEUDOSCALAR, coefficient},
              {$SCALAR, coefficient}},
  ... }; 
\end{lstlisting}
Here, {\tt \$VEV, \$PSEUDOSCALAR} and {\tt \$SCALAR} are the names chosen for the different components. 
Hence, for the two neutral Higgs doublets after electroweak symmetry breaking (EWSB) this reads
\begin{lstlisting}
DEFINITION[EWSB][VEVs]= 
  {{SHd0, {vd, 1/Sqrt[2]}, {sigmad,I/Sqrt[2]},{phid,1/Sqrt[2]}},
   {SHu0, {vu, 1/Sqrt[2]}, {sigmau,I/Sqrt[2]},{phiu,1/Sqrt[2]}}}; 
\end{lstlisting}
Two new possibilities have been added to the definition of VEVs in
\SARAHv{4}: the VEVs can be aligned, i.e. it is possible that only specific generations of a 
scalar field receive a VEV and the real and imaginary parts of the VEVs can be defined 
separately. Both 
changes are especially useful for the \Vevacious output presented in \ref{sec:vevacious}. 
\subsubsection{Aligned VEVs}
The standard definition of a model with broken electric charge due to charged slepton VEVs looks like
\begin{lstlisting}
DEFINITION[EWSB][VEVs]= 
  {...
   {SeL, {vL, 1/Sqrt[2]}, {sigmaL,I/Sqrt[2]},{phiL,1/Sqrt[2]}},
   {SeR, {vR, 1/Sqrt[2]}, {sigmaR,I/Sqrt[2]},{phiR,1/Sqrt[2]}},
  };
\end{lstlisting}
With this definition, all three generations of left and right sleptons would get a VEV. However, 
usually one is only interested in the case that staus receive VEVs. To restrict the possibility 
to obtain a VEV to the third generation the following syntax is used
\begin{lstlisting}
DEFINITION[EWSB][VEVs]= 
  {..,
   {SeL, {vL[3], 1/Sqrt[2]}, {sigmaL,I/Sqrt[2]},{phiL,1/Sqrt[2]}},
   {SeR, {vR[3], 1/Sqrt[2]}, {sigmaR,I/Sqrt[2]},{phiR,1/Sqrt[2]}}};
\end{lstlisting}
If one wants to consider smuon and stau VEVs, {\tt vL[2,3]}, {\tt vR[2,3]} can be used. 

\subsubsection{Complex VEVs}
To define complex VEVs, it has been possible in previous
versions of \SARAH to give the phase as last argument:
\begin{lstlisting}
DEFINITION[EWSB][VEVs]= 
  {{SHd0, {vd, 1/Sqrt[2]}, {sigmad,I/Sqrt[2]},{phid,1/Sqrt[2]}},
   {SHu0, {vu, 1/Sqrt[2]}, {sigmau,I/Sqrt[2]},{phiu,1/Sqrt[2]},{eta}}}; 
\end{lstlisting}
This is understood as $H_u^0 \to \frac{\exp(i \eta)}{\sqrt{2}} \left(v_u + i \sigma_u + \phi_u\right)$. In \SARAHv{4}, 
another possibility to define complex VEVs is to use 
\begin{lstlisting}
DEFINITION[EWSB][VEVs]= 
 {{SHd0, {vdR, 1/Sqrt[2]}, {vdI, I/Sqrt[2]},  
                      {sigmad,I/Sqrt[2]},{phid,1/Sqrt[2]}},
  {SHu0, {vuR, 1/Sqrt[2]}, {vuI, I/Sqrt[2]}, 
                      {sigmau,I/Sqrt[2]},{phiu,1/Sqrt[2]}}
   }; 
\end{lstlisting}
which is understood as
\begin{equation}
H_d^0 \to \frac{1}{\sqrt{2}}\left(v_d^R + i v_d^I + i \sigma_d + \phi_d \right)\,,\hspace{1cm} 
H_u^0 \to \frac{1}{\sqrt{2}}\left(v_u^R + i v_u^I + i \sigma_u + \phi_u \right) \, .
\end{equation}
This format has the advantage that the tree-level tadpole equations are also in the complex 
case polynomials in the VEVs. Therefore, they can be solved numerically with dedicated codes like {\tt HOM4PS2} \cite{lee2008hom4ps} 
which is used by \Vevacious.

\subsection{Numerical Solutions of Tadpole Equations in \SPheno}
There has been so far a major restriction in the definition of the properties of the \SPheno code: the tadpole 
equations have been solved on the \SARAH side using the {\tt Solve} command of \Mathematica. Because of this
it was necessary to choose a set of parameters for which an analytical solution of the tadpole equations exists. 
In the new version, this constrain is no longer present, but the tadpole equations can be solved numerically 
during the run of \SPheno. The numerical solution is based on a Broydn method and finds the solution which 
is the closest one to a given starting point. This option needs at least \SPheno 3.2.4. \\

We demonstrate the possibility to create \SPheno modules using a numerical solution of the 
tadpole equations at the example of the MSSM. 
To use a numerical solution and to define the starting point, {\tt SPheno.m} has to contain the following entries:
\lstset{frame=shadowbox}
\begin{lstlisting}
ParametersToSolveTadpoles = {\[Mu],B[\[Mu]]};
NumericalSolutionTadpoleEquations = True;
InitializationTadpoleParameters = { \[Mu] -> m0, B[\[Mu]]-> m0^2};
\end{lstlisting}
The first line is the same as for the analytical approach and defines that the tadpole equations have to be solved with respect to
$\mu$ and $B_\mu$. Without the other two lines, {\tt Mathematica's} function {\tt Solve}  would be used for this and the analytical solution 
would be exported to the Fortran code. However, due to the second line the attempts to solve the tadpole equations 
inside \Mathematica are skipped. The third line assumes that $\mu$ is $O(m_0)$ and $B_\mu$ is $O(m_0^2)$ 
($m_0$ is the unified scalar mass at the GUT scale) and the numerical routines use those values as initialization. 
Of course, other possible and reasonable choices would have been to relate $\mu, B_\mu$ with the running 
soft-breaking terms of the Higgs
\begin{lstlisting}
InitializationTadpoleParameters = { \[Mu] -> Sqrt[mHd2], B[\[Mu]]-> mHd2};
\end{lstlisting}
or to use fixed values
\begin{lstlisting}
InitializationTadpoleParameters = { \[Mu] -> 10^3, B[\[Mu]]-> 10^6};
\end{lstlisting}
Usually, the time needed to find the solution changes only slightly with the chosen initialization values
as long as they are not completely off. 

Note, all choices above would only find the solution with positive
$\mu$. Since an analytical solution, if it exists, contains all possible extrema, this is usually the preferred
method. 
Moreover, one has to be careful because the numerical approach only finds one minimum which must not be the 
global one. This could especially happen if one solves with respect to some VEVs which enter the equations
with third power. To check if the found minimum is really the global one or to calculate the life time of a
metastable point one can use  \Vevacious. 

\subsection{Running VEVs including Gauge Dependence}
\label{app:RunningVEVs}
\SARAHv{4} does not only calculate the one- and two-loop $\beta$-functions for all VEVs present in the model, 
but it includes also the gauge dependence of these RGEs. This is done by including the results of Ref.~\cite{Sperling:2013eva}.
To express the gauge dependence a variable {\tt Xi} is introduced and {\tt Xi} equal to 0 corresponds to Landau and equal to 1 
to Feynman gauge. With this convention the one- and two-loop $\beta$-function for $v_d$ in the MSSM reads now ({\tt BetaVEV[[1]]})
\begin{lstlisting}
{vd, 
  (vd*(3*(g1^2 + 5*g2^2)*(1 + Xi) - 60*trace[Yd, Adj[Yd]] 
     - 20*trace[Ye, Adj[Ye]]))/20, 
 -(vd*(207*g1^4 + 90*g1^2*g2^2 + 600*g2^4 + 500*g2^4*Xi + 75*g2^4*Xi^2 + 
      20*(g1^2*(-4 + 9*Xi) + 5*(32*g3^2 + 9*g2^2*Xi))*trace[Yd, Adj[Yd]] + 
      60*(5*g2^2*Xi + g1^2*(4 + Xi))*trace[Ye, Adj[Ye]] - 
      1800*trace[Yd, Adj[Yd], Yd, Adj[Yd]] - 
      600*trace[Yd, Adj[Yu], Yu, Adj[Yd]] - 
      600*trace[Ye, Adj[Ye], Ye, Adj[Ye]]))/200}
\end{lstlisting}

\subsection{Loop Corrections}
\label{sec:loops}
\SARAH calculates the one-loop corrections to the 1- and 2-point 
functions which can be used for a calculation of the one-loop corrected 
mass spectrum. This is done in 't Hooft gauge and so far only the 
$\overline{\text{DR}}$ scheme has been used. However,  it is more convenient to 
perform these calculations for non-SUSY models in the $\overline{\text{MS}}$
scheme. The results between both scheme differ by a constant term which is now
reflected in the variable {\tt rMS} introduced in \SARAHv{4}. 
{\tt rMS = 0} gives the results  in
 $\overline{\text{DR}}$ scheme and {\tt rMS = 1} corresponds to  $\overline{\text{MS}}$ scheme.

\lstset{frame=none}
\section{Summary of Commands}
\label{app:commands}
We only list here the most important commands with their main options available in \SARAH. 
For more details we refer to the manual. 
\subsection{Tree Level Relations}
\begin{itemize}
 \item Mass matrices \\
 The mass matrix for a particle ({\tt Particle})  is returned by
\begin{lstlisting}
 MassMatrix[Particle]
\end{lstlisting}
\item Tadpole equations \\
The tadpole equation corresponding to a scalar ({\tt Scalar}) is printed by using
\begin{lstlisting}
 TadpoleEquation[Scalar]
\end{lstlisting}
\item Vertices \\
To calculate the vertices for a list of external states ({\tt Particles}) use
\begin{lstlisting}
 Vertex[{Particles},Options];
\end{lstlisting}
The options define the considered eigenstates ({\tt Eigenstates -> EWSB/GaugeES}) as well as the treatment of 
dependences among parameters ({\tt UseDependences -> False/True}). \\
All vertices for a set of eigenstates are calculated at once  by
\begin{lstlisting}
MakeVertexList[$EIGENSTATES, Options];
\end{lstlisting}
Here, first the eigenstates ({\tt \$EIGENSTATES: GaugeES, EWSB}) are defined and as option it can be defined if only specific generic 
classes should be considered (e.g. {\tt GenericClasses -> FFS}). 
\end{itemize}

\subsection{Loop Calculations}
\begin{itemize}
 \item Renormalization Group Equations\\
The calculation of the RGEs at the one- and two-loop level can be started after the initialization of a model via
\begin{lstlisting}
CalcRGEs[Options];
\end{lstlisting}
See sec.~\ref{sec:RGEs} for the different options and the output for non-SUSY models. The output for SUSY models is 
saved in the following arrays:
\begin{itemize}
\item \verb"Gij": Anomalous dimensions of all chiral superfields
\item \verb"BetaWijkl": Quartic superpotential parameters
\item \verb"BetaYijk": Trilinear superpotential parameters
\item \verb"BetaMuij": Bilinear superpotential parameters
\item \verb"BetaLi": Linear superpotential parameters
\item \verb"BetaQijkl": Quartic soft-breaking parameters
\item \verb"BetaTijk": Trilinear soft-breaking parameters
\item \verb"BetaBij": Bilinear soft-breaking parameters
\item \verb"BetaSLi": Linear soft-breaking parameters
\item \verb"Betam2ij": Scalar squared masses
\item \verb"BetaMi": Majorana Gaugino masses
\item \verb"BetaGauge": Gauge couplings
\item \verb"BetaVEVs": VEVs
\item \verb"BetaDGi": Dirac gaugino mass terms
\end{itemize}
\item One-loop Tadpoles and Self-Energies\\
Loop corrections are calculated via
\begin{lstlisting}
CalcLoopCorrections[$EIGENSTATES]; 
\end{lstlisting}
As argument only the considered eigenstates (e.g. {\tt \$EIGENSTATES=EWSB}) have to be defined. The results are saved in
the variables {\tt Tadpoles1LoopSums[\$EIGENSTATES]} and {\tt SelfEnergy1LoopSum[\$EIGENSTATES]}
as sums of all contributions, or as list of the different contributions  in 
{\tt Tadpoles1LoopList[\$EIGENSTATES]} and  {\tt SelfEnergy1LoopList[\$EIGENSTATES]}.
\end{itemize}

\subsection{Model Files and \LaTeX\ Output}
\begin{itemize}
 \item \UFO format\\
 Model files in the \UFO format are written by
\begin{lstlisting}
MakeUFO[Options]
\end{lstlisting}
As option it can be defined, if specific generic classes (like four scalar interactions) 
should be excluded in the output (e.g. {\tt Exclude -> \{SSSS\}}).
\item \CalcHep/\CompHep  \\
Model files for \CalcHep which also can be used with \MO are generated by
\begin{lstlisting}
MakeCHep[Options]
\end{lstlisting}
The options define for instance if the \CompHep format should be used instead of the 
\CalcHep format ({\tt CompHep->False/True}), 
if Feynman gauge and/or CP violation should be included ({\tt FeynmanGauge -> True/False}, 
{\tt CPViolation -> True/False}) and how the 
masses of fields should be obtained by \CalcHep ({\tt SLHAinput}, {\tt CalculateMasses}, {\tt RunSPhenoViaCalcHep}). 
\item \WHIZARD\\
The model files for \WHIZARD and \OMEGA are generated by
\begin{lstlisting}
MakeWHIZARD[Options]
\end{lstlisting}
The options can be used to define the generic classes of vertices which 
should be excluded (e.g. {\tt Exclude -> \{SSSS\}}), and the maximal number of
couplings per file ({\tt MaximalCouplingsPerFile -> \#}). 
\item \FeynArts/\FormCalc
Model files for \FeynArts are written by 
\begin{lstlisting}
MakeFeynArts[Options]
\end{lstlisting}
As option the treatment of counter terms can be defined ({\tt AddCounterTerms -> True/False})
\item \Vevacious
The model files for \Vevacious are written by
\begin{lstlisting}
MakeVevacious[Options]
\end{lstlisting}
For the different options, see sec.~\ref{sec:vevacious}. 
\item \SPheno \\
To output the \SPheno code, the command
\begin{lstlisting}
MakeSPheno[Options];
\end{lstlisting}
has to be used. The options define the file which contains the 
necessary input for a custom made \SPheno module ({\tt InputFile -> \$FILENAME}), defines
the standard compiler ({\tt StandardCompiler -> \$COMPILER}), and if the results from
previous calculations should be used  ({\tt ReadLists -> True/False})
\item \LaTeX:\\
All derived information (mass matrices, vertices, RGEs, loop corrections) can be exported to 
\LaTeX\ to get the expressions in a human readable form. This is done via
\begin{lstlisting}
MakeTeX[Options] 
\end{lstlisting}
The options define if Feynman diagrams should be painted with {\tt FeynMF} \cite{Ohl:FeynMF} 
for all vertices \newline({\tt FeynmanDiagrams -> True/False}), 
and if \SARAH specific information (e.g. internal names for fields and parameters) should be 
written as well ({\tt WriteSARAH -> True/False}). 
\end{itemize}

\section{Models}
\label{app:models}
\subsection{Supersymmetric Models}
\begin{itemize}
 \item Minimal supersymmetric standard model 
    \begin{itemize} 
     \item With general flavor and CP structure ({\tt MSSM})
     \item Without flavor violation ({\tt MSSM/NoFV})
     \item With explicit CP violation in the Higgs sector ({\tt MSSM/CPV})
     \item In SCKM basis ({\tt MSSM/CKM})
    \end{itemize}
   \item Singlet extensions: 
   \begin{itemize}
    \item Next-to-minimal supersymmetric standard model ({\tt NMSSM}, {\tt NMSSM/NoFV}, {\tt NMSSM/CPV}, {\tt NMSSM/CKM})
    \item near-to-minimal supersymmetric standard model  ({\tt near-MSSM}) 
    \item General singlet extended, supersymmetric standard model ({\tt SMSSM})  
  \end{itemize}
  \item Triplet extensions 
  \begin{itemize} 
    \item Triplet extended MSSM ({\tt TMSSM}) 
    \item Triplet extended NMSSM ({\tt TNMSSM}) 
  \end{itemize}
   \item Models with $R$-parity violation  
  \begin{itemize}
    \item bilinear RpV ({\tt MSSM-RpV/Bi}) 
    \item Lepton number violation ({\tt MSSM-RpV/LnV})
    \item Only trilinear lepton number violation ({\tt MSSM-RpV/TriLnV})
    \item Baryon number violation ({\tt MSSM-RpV/BnV})  
    \item $\mu\nu$SSM ({\tt munuSSM}) 
  \end{itemize}
   \item Additional $U(1)'s$ 
  \begin{itemize}
    \item $U(1)$-extended MSSM ({\tt UMSSM})  
    \item secluded MSSM ({\tt secluded-MSSM}) 
    \item minimal $B-L$ model ({\tt B-L-SSM})  
    \item minimal singlet-extended $B-L$ model ({\tt N-B-L-SSM})
  \end{itemize}
   \item SUSY-scale seesaw extensions
    \begin{itemize}
      \item inverse seesaw ({\tt inverse-Seesaw}) 
      \item linear seesaw ({\tt LinSeesaw}) 
      \item singlet extended inverse seesaw ({\tt inverse-Seesaw-NMSSM}) 
      \item inverse seesaw with $B-L$ gauge group ({\tt B-L-SSM-IS})  
      \item minimal $U(1)_R \times U(1)_{B-L}$ model with inverse seesaw  ({\tt BLRinvSeesaw}) 
\end{itemize}
 \item Models with Dirac Gauginos
   \begin{itemize}
    \item MSSM/NMSSM with Dirac Gauginos ({\tt DiracGauginos}) 
    \item minimal $R$-Symmetric SSM ({\tt MRSSM}) 
   \end{itemize}
 \item High-scale extensions
\begin{itemize}
 \item Seesaw 1 - 3 ($SU(5)$ version) ,
 ({\tt Seesaw1},{\tt Seesaw2},{\tt Seesaw3}) 
 \item Left/right model ($\Omega$LR) ({\tt Omega})
\end{itemize}
\end{itemize}

\subsection{Non-Supersymmetric Models}
\begin{itemize}
\item Standard Model (SM) ({\tt SM}), Standard model in CKM basis ({\tt SM/CKM}) 
\item inert Higgs doublet model ({\tt Inert}) 
\item B-L extended SM ({\tt B-L-SM})
\item B-L extended SM with inverse seesaw ({\tt B-L-SM-IS})
\item SM extended by a scalar color octet ({\tt SM-8C})
\item Two Higgs doublet model ({\tt THDM})
\item Singlet extended SM ({\tt SSM})
\end{itemize}

\lstset{basicstyle=\scriptsize, frame=shadowbox}
\section{The Model File of the NMSSM in \SARAHv{4}}
\label{app:NMSSM}
\begin{lstlisting}
(*-------------------------------------------*)
(* Global symmetries *)
(*-------------------------------------------*)

Global[[1]]={Z[2],RParity}; 
Global[[2]]={Z[3],Z3}; 
RpM = {-1,-1,1}; RpP = {1,1,-1};

(*-------------------------------------------*)
(*   Particle Content*)
(*-------------------------------------------*)

(* Vector Superfields *)

Gauge[[1]]={B,   U[1], hypercharge, g1,False, RpM, 0};
Gauge[[2]]={WB, SU[2], left,        g2,True,  RpM, 0};
Gauge[[3]]={G,  SU[3], color,       g3,False, RpM, 0};

(* Chiral Superfields *)

SuperFields[[1]] = {q, 3, {uL,  dL},    1/6, 2, 3, RpM, 1/3};  
SuperFields[[2]] = {l, 3, {vL,  eL},   -1/2, 2, 1, RpM, 1/3};
SuperFields[[3]] = {Hd,1, {Hd0, Hdm},  -1/2, 2, 1, RpP, 1/3};
SuperFields[[4]] = {Hu,1, {Hup, Hu0},   1/2, 2, 1, RpP, 1/3};

SuperFields[[5]] = {d, 3, conj[dR],   1/3, 1, -3, RpM, 1/3};
SuperFields[[6]] = {u, 3, conj[uR],  -2/3, 1, -3, RpM, 1/3};
SuperFields[[7]] = {e, 3, conj[eR],     1, 1,  1, RpM, 1/3};

SuperFields[[8]] = {s, 1, sR,     0, 1,  1, RpP, 1/3};


(*------------------------------------------------------*)
(* Superpotential *)
(*------------------------------------------------------*)

SuperPotential = Yu u.q.Hu - Yd d.q.Hd - Ye e.l.Hd + \[Lambda] Hu.Hd.s + \[Kappa]/3 s.s.s;


(*----------------------------------------------*)
(*   ROTATIONS                                  *)
(*----------------------------------------------*)

NameOfStates={GaugeES, EWSB};

(* ----- After EWSB ----- *)

DEFINITION[EWSB][GaugeSector] =
{ 
  {{VB,VWB[3]},{VP,VZ},ZZ},
  {{VWB[1],VWB[2]},{VWm,conj[VWm]},ZW},
  {{fWB[1],fWB[2],fWB[3]},{fWm,fWp,fW0},ZfW}
};      
        
 
DEFINITION[EWSB][VEVs]= 
{    {SHd0, {vd, 1/Sqrt[2]}, {sigmad,I/Sqrt[2]},{phid,1/Sqrt[2]}},
     {SHu0, {vu, 1/Sqrt[2]}, {sigmau,I/Sqrt[2]},{phiu,1/Sqrt[2]}},
     {SsR, {vS, 1/Sqrt[2]}, {sigmaS,I/Sqrt[2]},{phiS,1/Sqrt[2]}}     };


 
DEFINITION[EWSB][MatterSector]= 

{    {{SdL, SdR}, {Sd, ZD}},
     {{SvL}, {Sv, ZV}},
     {{SuL, SuR}, {Su, ZU}},
     {{SeL, SeR}, {Se, ZE}},
     {{phid, phiu, phiS}, {hh, ZH}},
     {{sigmad, sigmau,sigmaS}, {Ah, ZA}},
     {{SHdm,conj[SHup]},{Hpm,ZP}},
     {{fB, fW0, FHd0, FHu0,FsR}, {L0, ZN}}, 
     {{{fWm, FHdm}, {fWp, FHup}}, {{Lm,UM}, {Lp,UP}}},
     {{{FeL},{conj[FeR]}},{{FEL,ZEL},{FER,ZER}}},
     {{{FdL},{conj[FdR]}},{{FDL,ZDL},{FDR,ZDR}}},
     {{{FuL},{conj[FuR]}},{{FUL,ZUL},{FUR,ZUR}}}            
   }; 

DEFINITION[EWSB][Phases]= 
{    {fG, PhaseGlu}
    }; 

DEFINITION[EWSB][DiracSpinors]={
 Fd ->{  FDL, conj[FDR]},
 Fe ->{  FEL, conj[FER]},
 Fu ->{  FUL, conj[FUR]},
 Fv ->{  FvL, 0},
 Chi ->{ L0, conj[L0]},
 Cha ->{ Lm, conj[Lp]},
 Glu ->{ fG, conj[fG]}
}; 
\end{lstlisting}

\section{The Model File of the SM in \SARAHv{4}}
\label{app:SM}
\begin{lstlisting}
(*-------------------------------------------*)
(*   Particle Content*)
(*-------------------------------------------*)

(* Gauge Groups *)

Gauge[[1]]={B,   U[1], hypercharge, g1,False};
Gauge[[2]]={WB, SU[2], left,        g2,True};
Gauge[[3]]={G,  SU[3], color,       g3,False};


(* Matter Fields *)

FermionFields[[1]] = {q, 3, {uL, dL},     1/6, 2,  3};  
FermionFields[[2]] = {l, 3, {vL, eL},    -1/2, 2,  1};
FermionFields[[3]] = {d, 3, conj[dR],     1/3, 1, -3};
FermionFields[[4]] = {u, 3, conj[uR],    -2/3, 1, -3};
FermionFields[[5]] = {e, 3, conj[eR],       1, 1,  1};

ScalarFields[[1]] =  {H, 1, {Hp, H0},     1/2, 2,  1};


        
(*----------------------------------------------*)
(*   DEFINITION                                 *)
(*----------------------------------------------*)

NameOfStates={GaugeES, EWSB};

(* ----- Potential ----- *)


LagNoHC = -Mu2 conj[H].H + 1/2 \[Lambda] conj[H].H.conj[H].H;
LagHC =  Yd conj[H].d.q + Ye conj[H].e.l + Yu H.u.q;

DEFINITION[GaugeES][LagrangianInput]= {
	{LagHC, {AddHC->True}},
	{LagNoHC,{AddHC->False}}
};
			  		  

(* Gauge Sector *)

DEFINITION[EWSB][GaugeSector] =
{ 
  {{VB,VWB[3]},{VP,VZ},ZZ},
  {{VWB[1],VWB[2]},{VWp,conj[VWp]},ZW}
};     
        
(* ----- VEVs ---- *)

DEFINITION[EWSB][VEVs]= 
{    {H0, {v, 1/Sqrt[2]}, {Ah,I/Sqrt[2]},{hh, 1/Sqrt[2]}}     };
 

DEFINITION[EWSB][MatterSector]=   
    {{{{dL}, {conj[dR]}}, {{DL,Vd}, {DR,Ud}}},
     {{{uL}, {conj[uR]}}, {{UL,Vu}, {UR,Uu}}},
     {{{eL}, {conj[eR]}}, {{EL,Ve}, {ER,Ue}}}};  


(*------------------------------------------------------*)
(* Dirac-Spinors *)
(*------------------------------------------------------*)

DEFINITION[EWSB][DiracSpinors]={
 Fd ->{  DL, conj[DR]},
 Fe ->{  EL, conj[ER]},
 Fu ->{  UL, conj[UR]},
 Fv ->{  vL, 0}};
\end{lstlisting}

\end{appendix}

\bibliography{lit.bib}

\begin{thebibliography}{10}

\bibitem{Staub:2008uz}
F.~Staub, ``{SARAH},'' 2008.

\bibitem{Staub:2009bi}
F.~Staub, ``{From Superpotential to Model Files for FeynArts and
  CalcHep/CompHep},'' {\em Comput.Phys.Commun.}, vol.~181, pp.~1077--1086,
  2010.

\bibitem{Staub:2010jh}
F.~Staub, ``{Automatic Calculation of supersymmetric Renormalization Group
  Equations and Self Energies},'' {\em Comput.Phys.Commun.}, vol.~182,
  pp.~808--833, 2011.

\bibitem{Staub:2012pb}
F.~Staub, ``{SARAH 3.2: Dirac Gauginos, UFO output, and more},'' {\em Computer
  Physics Communications}, vol.~184, pp.~pp. 1792--1809, 2013.

\bibitem{Ender:2011qh}
K.~Ender, T.~Graf, M.~Muhlleitner, and H.~Rzehak, ``{Analysis of the NMSSM
  Higgs Boson Masses at One-Loop Level},'' {\em Phys.Rev.}, vol.~D85,
  p.~075024, 2012.

\bibitem{Graf:2012hh}
T.~Graf, R.~Grober, M.~Muhlleitner, H.~Rzehak, and K.~Walz, ``{Higgs Boson
  Masses in the Complex NMSSM at One-Loop Level},'' {\em JHEP}, vol.~1210,
  p.~122, 2012.

\bibitem{Kaminska:2013mya}
A.~Kaminska, G.~G. Ross, and K.~Schmidt-Hoberg, ``{Non-universal gaugino masses
  and fine tuning implications for SUSY searches in the MSSM and the GNMSSM},''
  2013.

\bibitem{List:2013dga}
J.~List and B.~Vormwald, ``{Bilinear R Parity Violation at the ILC - Neutrino
  Physics at Colliders},'' 2013.

\bibitem{Dreiner:2013jta}
H.~Dreiner, K.~Nickel, and F.~Staub, ``{$B_{s,d}^0 \to \mu\bar{\mu}$ and $B\to
  X_s\gamma$ in the R-parity violating MSSM},'' 2013.

\bibitem{Abada:2011mg}
A.~Abada, A.~Figueiredo, J.~Romao, and A.~Teixeira, ``{Probing the
  supersymmetric type III seesaw: LFV at low-energies and at the LHC},'' {\em
  JHEP}, vol.~1108, p.~099, 2011.

\bibitem{BhupalDev:2012ru}
P.~Bhupal~Dev, S.~Mondal, B.~Mukhopadhyaya, and S.~Roy, ``{Phenomenology of
  Light Sneutrino Dark Matter in cMSSM/mSUGRA with Inverse Seesaw},'' {\em
  JHEP}, vol.~1209, p.~110, 2012.

\bibitem{DeRomeri:2012qd}
V.~De~Romeri and M.~Hirsch, ``{Sneutrino Dark Matter in Low-scale Seesaw
  Scenarios},'' {\em JHEP}, vol.~1212, p.~106, 2012.

\bibitem{Esteves:2010si}
J.~Esteves, J.~Romao, M.~Hirsch, A.~Vicente, W.~Porod, {\em et~al.}, ``{LHC and
  lepton flavour violation phenomenology of a left-right extension of the
  MSSM},'' {\em JHEP}, vol.~1012, p.~077, 2010.

\bibitem{Krauss:2013jva}
M.~E. Krauss, W.~Porod, and F.~Staub, ``{SO(10) inspired gauge-mediated
  supersymmetry breaking},'' {\em Phys.Rev.}, vol.~D88, p.~015014, 2013.

\bibitem{O'Leary:2011yq}
B.~O'Leary, W.~Porod, and F.~Staub, ``{Mass spectrum of the minimal SUSY B-L
  model},'' {\em JHEP}, vol.~1205, p.~042, 2012.

\bibitem{Hirsch:2012kv}
M.~Hirsch, W.~Porod, L.~Reichert, and F.~Staub, ``{Phenomenology of the minimal
  supersymmetric $U(1)_{B-L}\times U(1)_R$ extension of the standard model},''
  {\em Phys.Rev.}, vol.~D86, p.~093018, 2012.

\bibitem{Frugiuele:2012pe}
C.~Frugiuele, T.~Gregoire, P.~Kumar, and E.~Ponton, ``{'L=R' - $U(1)_R$ as the
  Origin of Leptonic 'RPV'},'' {\em JHEP}, vol.~1303, p.~156, 2013.

\bibitem{Benakli:2012cy}
K.~Benakli, M.~D. Goodsell, and F.~Staub, ``{Dirac Gauginos and the 125 GeV
  Higgs},'' {\em JHEP}, vol.~1306, p.~073, 2013.

\bibitem{Alves:2012fx}
D.~S. Alves, P.~J. Fox, and N.~Weiner, ``{Supersymmetry with a Sister Higgs},''
  2012.

\bibitem{Pukhov:2004ca}
A.~Pukhov, ``{CalcHEP 2.3: MSSM, structure functions, event generation, batchs,
  and generation of matrix elements for other packages},'' 2004.

\bibitem{Boos:1994xb}
E.~Boos, M.~Dubinin, V.~Ilyin, A.~Pukhov, and V.~Savrin, ``{CompHEP:
  Specialized package for automatic calculations of elementary particle decays
  and collisions},'' 1994.

\bibitem{Hahn:2000kx}
T.~Hahn, ``{Generating Feynman diagrams and amplitudes with FeynArts 3},'' {\em
  Comput.Phys.Commun.}, vol.~140, pp.~418--431, 2001.

\bibitem{Hahn:2009bf}
T.~Hahn, ``{FormCalc 6},'' {\em PoS}, vol.~ACAT08, p.~121, 2008.

\bibitem{Kilian:2007gr}
W.~Kilian, T.~Ohl, and J.~Reuter, ``{WHIZARD: Simulating Multi-Particle
  Processes at LHC and ILC},'' {\em Eur.Phys.J.}, vol.~C71, p.~1742, 2011.

\bibitem{Moretti:2001zz}
M.~Moretti, T.~Ohl, and J.~Reuter, ``{O'Mega: An Optimizing matrix element
  generator},'' 2001.

\bibitem{Degrande:2011ua}
C.~Degrande, C.~Duhr, B.~Fuks, D.~Grellscheid, O.~Mattelaer, {\em et~al.},
  ``{UFO - The Universal FeynRules Output},'' {\em Comput.Phys.Commun.},
  vol.~183, pp.~1201--1214, 2012.

\bibitem{Alwall:2011uj}
J.~Alwall, M.~Herquet, F.~Maltoni, O.~Mattelaer, and T.~Stelzer, ``{MadGraph 5
  : Going Beyond},'' {\em JHEP}, vol.~1106, p.~128, 2011.

\bibitem{Cullen:2011ac}
G.~Cullen, N.~Greiner, G.~Heinrich, G.~Luisoni, P.~Mastrolia, {\em et~al.},
  ``{Automated One-Loop Calculations with GoSam},'' {\em Eur.Phys.J.},
  vol.~C72, p.~1889, 2012.

\bibitem{Gieseke:2003hm}
S.~Gieseke, A.~Ribon, M.~H. Seymour, P.~Stephens, and B.~Webber, ``{Herwig++
  1.0: An Event generator for e+ e- annihilation},'' {\em JHEP}, vol.~0402,
  p.~005, 2004.

\bibitem{Gieseke:2006ga}
S.~Gieseke, D.~Grellscheid, K.~Hamilton, A.~Ribon, P.~Richardson, {\em et~al.},
  ``{Herwig++ 2.0 Release Note},'' 2006.

\bibitem{Bellm:2013lba}
J.~Bellm, S.~Gieseke, D.~Grellscheid, A.~Papaefstathiou, S.~Platzer, {\em
  et~al.}, ``{Herwig++ 2.7 Release Note},'' 2013.

\bibitem{Porod:2003um}
W.~Porod, ``{SPheno, a program for calculating supersymmetric spectra, SUSY
  particle decays and SUSY particle production at e+ e- colliders},'' {\em
  Comput.Phys.Commun.}, vol.~153, pp.~275--315, 2003.

\bibitem{Porod:2011nf}
W.~Porod and F.~Staub, ``{SPheno 3.1: Extensions including flavour, CP-phases
  and models beyond the MSSM},'' {\em Comput.Phys.Commun.}, vol.~183,
  pp.~2458--2469, 2012.

\bibitem{Bechtle:2008jh}
P.~Bechtle, O.~Brein, S.~Heinemeyer, G.~Weiglein, and K.~E. Williams,
  ``{HiggsBounds: Confronting Arbitrary Higgs Sectors with Exclusion Bounds
  from LEP and the Tevatron},'' {\em Comput.Phys.Commun.}, vol.~181,
  pp.~138--167, 2010.

\bibitem{Bechtle:2011sb}
P.~Bechtle, O.~Brein, S.~Heinemeyer, G.~Weiglein, and K.~E. Williams,
  ``{HiggsBounds 2.0.0: Confronting Neutral and Charged Higgs Sector
  Predictions with Exclusion Bounds from LEP and the Tevatron},'' {\em
  Comput.Phys.Commun.}, vol.~182, pp.~2605--2631, 2011.

\bibitem{Bechtle:2013xfa}
P.~Bechtle, S.~Heinemeyer, O.~Stål, T.~Stefaniak, and G.~Weiglein,
  ``{HiggsSignals: Confronting arbitrary Higgs sectors with measurements at the
  Tevatron and the LHC},'' 2013.

\bibitem{Fonseca:2011sy}
R.~M. Fonseca, ``{Calculating the renormalisation group equations of a SUSY
  model with Susyno},'' {\em Comput.Phys.Commun.}, vol.~183, pp.~2298--2306,
  2012.

\bibitem{Camargo-Molina:2013qva}
J.~Camargo-Molina, B.~O'Leary, W.~Porod, and F.~Staub, ``{Vevacious: A Tool For
  Finding The Global Minima Of One-Loop Effective Potentials With Many
  Scalars},'' 2013.

\bibitem{Wainwright:2011kj}
C.~L. Wainwright, ``{CosmoTransitions: Computing Cosmological Phase Transition
  Temperatures and Bubble Profiles with Multiple Fields},'' {\em
  Comput.Phys.Commun.}, vol.~183, pp.~2006--2013, 2012.

\bibitem{Semenov:1998eb}
A.~Semenov, ``{LanHEP: A package for automatic generation of Feynman rules from
  the Lagrangian},'' {\em Comput.Phys.Commun.}, vol.~115, pp.~124--139, 1998.

\bibitem{Semenov:2002jw}
A.~Semenov, ``{LanHEP: A Package for automatic generation of Feynman rules in
  field theory. Version 2.0},'' 2002.

\bibitem{Semenov:2008jy}
A.~Semenov, ``{LanHEP: A Package for the automatic generation of Feynman rules
  in field theory. Version 3.0},'' {\em Comput.Phys.Commun.}, vol.~180,
  pp.~431--454, 2009.

\bibitem{Semenov:2010qt}
A.~Semenov, ``{LanHEP - a package for automatic generation of Feynman rules
  from the Lagrangian. Updated version 3.1},'' 2010.

\bibitem{Christensen:2008py}
N.~D. Christensen and C.~Duhr, ``{FeynRules - Feynman rules made easy},'' {\em
  Comput.Phys.Commun.}, vol.~180, pp.~1614--1641, 2009.

\bibitem{Christensen:2010wz}
N.~D. Christensen, C.~Duhr, B.~Fuks, J.~Reuter, and C.~Speckner, ``{Introducing
  an interface between WHIZARD and FeynRules},'' {\em Eur.Phys.J.}, vol.~C72,
  p.~1990, 2012.

\bibitem{Bharucha:2013ela}
A.~Bharucha, A.~Goudelis, and M.~McGarrie, ``{En-gauging Naturalness},'' 2013.

\bibitem{Martin:1993zk}
S.~P. Martin and M.~T. Vaughn, ``{Two loop renormalization group equations for
  soft supersymmetry breaking couplings},'' {\em Phys.Rev.}, vol.~D50, p.~2282,
  1994.

\bibitem{Fonseca:2011vn}
R.~M. Fonseca, M.~Malinsky, W.~Porod, and F.~Staub, ``{Running soft parameters
  in SUSY models with multiple U(1) gauge factors},'' {\em Nucl.Phys.},
  vol.~B854, pp.~28--53, 2012.

\bibitem{Goodsell:2012fm}
M.~D. Goodsell, ``{Two-loop RGEs with Dirac gaugino masses},'' {\em JHEP},
  vol.~1301, p.~066, 2013.

\bibitem{Machacek:1983tz}
M.~E. Machacek and M.~T. Vaughn, ``{Two Loop Renormalization Group Equations in
  a General Quantum Field Theory. 1. Wave Function Renormalization},'' {\em
  Nucl. Phys.}, vol.~B222, p.~83, 1983.

\bibitem{Machacek:1983fi}
M.~E. Machacek and M.~T. Vaughn, ``{Two Loop Renormalization Group Equations in
  a General Quantum Field Theory. 2. Yukawa Couplings},'' {\em Nucl. Phys.},
  vol.~B236, p.~221, 1984.

\bibitem{Machacek:1984zw}
M.~E. Machacek and M.~T. Vaughn, ``{Two Loop Renormalization Group Equations in
  a General Quantum Field Theory. 3. Scalar Quartic Couplings},'' {\em Nucl.
  Phys.}, vol.~B249, p.~70, 1985.

\bibitem{Luo:2002ti}
M.-x. Luo, H.-w. Wang, and Y.~Xiao, ``{Two-loop renormalization group equations
  in general gauge field theories},'' {\em Phys. Rev.}, vol.~D67, p.~065019,
  2003.

\bibitem{Fonseca:2013bua}
R.~M. Fonseca, M.~Malinsky, and F.~Staub, ``{Renormalization group equations
  and matching in a general quantum field theory with kinetic mixing},'' 2013.

\bibitem{Sperling:2013eva}
M.~Sperling, D.~Stöckinger, and A.~Voigt, ``{Renormalization of vacuum
  expectation values in spontaneously broken gauge theories},'' {\em JHEP},
  vol.~1307, p.~132, 2013.

\bibitem{Lyonnet:2013dna}
F.~Lyonnet, I.~Schienbein, F.~Staub, and A.~Wingerter, ``{PyR@TE:
  Renormalization Group Equations for General Gauge Theories},'' 2013.

\bibitem{Dreiner:2012dh}
H.~Dreiner, K.~Nickel, W.~Porod, and F.~Staub, ``{Full 1-loop calculation of
  BR$(B_{s,d}^0\to \ell \bar \ell)$ in models beyond the MSSM with SARAH and
  SPheno},'' {\em Comput.Phys.Commun.}, vol.~184, pp.~2604--2617, 2013.

\bibitem{Staub:2011dp}
F.~Staub, T.~Ohl, W.~Porod, and C.~Speckner, ``{A Tool Box for Implementing
  Supersymmetric Models},'' {\em Comput.Phys.Commun.}, vol.~183,
  pp.~2165--2206, 2012.

\bibitem{Belanger:2010st}
G.~Belanger, N.~D. Christensen, A.~Pukhov, and A.~Semenov, ``{SLHAplus: a
  library for implementing extensions of the standard model},'' {\em
  Comput.Phys.Commun.}, vol.~182, pp.~763--774, 2011.

\bibitem{Camargo-Molina:2013sta}
J.~Camargo-Molina, B.~O'Leary, W.~Porod, and F.~Staub, ``{Stability of the
  CMSSM against sfermion VEVs},'' 2013.

\bibitem{Martin:2001vx}
S.~P. Martin, ``{Two loop effective potential for a general renormalizable
  theory and softly broken supersymmetry},'' {\em Phys.Rev.}, vol.~D65,
  p.~116003, 2002.

\bibitem{lee2008hom4ps}
T.~Lee, T.~Li, and C.~Tsai, ``Hom4ps-2.0: a software package for solving
  polynomial systems by the polyhedral homotopy continuation method,'' {\em
  Computing}, vol.~83, no.~2, pp.~109--133, 2008.

\bibitem{Ohl:FeynMF}
T.~Ohl, ``Feynmf: Drawing feynman diagrams with latex and metafont,'' 1997.

\end{thebibliography}
\bibliographystyle{ieeetr}

\end{document}